\documentclass[11pt]{article}
\usepackage{graphicx}
\def \d {{\rm d}}

\begin{document}

\title{An interpretation of Robinson--Trautman type N solutions} 

\author{J. B. Griffiths$^1$\thanks{E--mail: {\tt J.B.Griffiths@Lboro.ac.uk}}, \ 
J. Podolsk\'y$^2$\thanks{E--mail: {\tt Podolsky@mbox.troja.mff.cuni.cz}} \ and
P. Docherty$^1$  \\ \\ 
$^1$ Department of Mathematical Sciences, Loughborough University, \\ 
Loughborough, Leics. LE11 3TU, U.K. \\ \\
$^2$ Institute of Theoretical Physics, Charles University in Prague,\\
V Hole\v{s}ovi\v{c}k\'ach 2, 18000 Prague 8, Czech Republic.\\
\\}

\date{\today}
\maketitle

\begin{abstract}
\noindent
The Robinson--Trautman type~N solutions, which describe expanding gravitational
waves, are investigated for all possible values of the cosmological
constant~$\Lambda$ and the curvature parameter~$\epsilon$. The wave surfaces are
always (hemi-)spherical, with successive surfaces displaced in a way which depends
on~$\epsilon$. Explicit sandwich waves of this class are studied in Minkowski,
de~Sitter or anti-de~Sitter backgrounds. A particular family of such solutions
which can be used to represent snapping or decaying cosmic strings is considered
in detail, and its singularity and global structure is presented.  
\end{abstract}

\section{Introduction}

The family of Robinson--Trautman solutions \cite{RobTra60}--\cite{KSMH80} is well
known. They are algebraically special solutions having a repeated principal null
direction which is associated with an expanding, shear-free and twist-free null
geodesic congruence. However, in general, these solutions are still not well
understood. A complete physical interpretation is only known for a few very
special cases (of type~D). It is the purpose of the present paper to clarify
the physical meaning of the type~N solutions of this class.

These Robinson--Trautman type~N solutions can be classified \cite{GarPle81},
\cite{BicPod99a} as $RTN(\Lambda,\epsilon)$, where the two constant parameters are
the cosmological constant $\Lambda$, and the Gaussian curvature $2\epsilon$ of
certain privileged spatial sections. Some physical aspects of these solutions
have been investigated in \cite{BicPod99b} using an analysis of geodesic
deviation.

The approach to be adopted in the present paper is as follows: -- First we
consider the weak field limit in which the metric approaches the background
Minkowski, de~Sitter or anti-de~Sitter space-times. In this limit, it is possible
to determine exactly the geometrical properties of the wave surfaces and the
locations of the singularities which appear in the metric. After this, we
consider an explicit family of exact sandwich wave solutions. Since these
propagate into Minkowski, de~Sitter or anti-de~Sitter backgrounds, the geometry
of the wavefronts can again be analysed completely.

In particular, it is found that some sandwich wave solutions of this type may be
interpreted in terms of snapping and decaying cosmic strings in the corresponding
backgrounds. One simple case of this has been briefly presented \cite{GriDoc02},
but the full family of such solutions in different backgrounds and for different
families of wave surfaces and profiles is described here.

The sandwich wave solutions considered here can all be reduced to their impulsive
limits. These describe spherical impulsive gravitational waves in backgrounds of
constant curvature that are generated by snapping (or expanding) cosmic strings.
Such solutions are well known \cite{Penrose72}--\cite{PodGri01} and can be
constructed by various methods in all these backgrounds. It has been observed,
however, that these cannot be considered as impulsive Robinson--Trautman
solutions in a rigorous sense, as these would involve the square of the delta
function appearing in the metric when written in Garc\'{\i}a--Pleba\'nski
coordinates. Nevertheless, it has recently been argued \cite{PodGri99a} that
these limits are indeed reasonable. The impulsive limits of the sandwich waves
presented in \cite{GriDoc02} and below confirm this. Moreover, they provide a
framework in which these limits can be investigated more rigorously.

\section{The Robinson--Trautman type~N solutions}

All Robinson--Trautman solutions of Petrov type N, which are necessarily vacuum
but with a possibly non-zero cosmological constant $\Lambda$, are given by the
line element 
 \begin{equation}
 \d s^2 =2\,\d u\,\d r
+\Big[2\epsilon-2r(\log P)_u -{\Lambda\over3}r^2\Big]\d u^2
-2{r^2\over P^2}\,\d\zeta\,\d\bar\zeta, 
 \label{RTmetric}
 \end{equation} 
 where the null coordinate $u$ can be regarded as retarded time, $r$ is a
Bondi-type luminosity distance, and $\zeta$ is a complex stereographic-type
coordinate. The function $2\epsilon(u)$ is the Gaussian curvature of the
2-surfaces $2P^{-2}\d\zeta\d\bar\zeta$. A coordinate freedom can always be used
to set $\epsilon$ to $+1$, $0$ or $-1$. The function $P(\zeta,\bar\zeta,u)$ has to
satisfy the field equation \ $P^2(\log P)_{\zeta\bar\zeta}=\epsilon$, \ which has
the general solution \cite{FosNew67}
 \begin{equation}
 P=\big(1+\epsilon F\bar F\big) 
\big(F_\zeta\bar F_{\bar\zeta}\big)^{-1/2}, 
 \label{P}
 \end{equation} 
 where $F=F(\zeta,u)$ is an arbitrary complex function of $u$ and $\zeta$,
holomorphic in $\zeta$. Using a natural tetrad, the only non-zero component of
the Weyl tensor is given by 
 \begin{equation}
 \bar\Psi_4 =-{1\over r}\left[ P^2\big(\log P\big)_{u\zeta} \right]_{\zeta}
 ={P^2\over2r}F_\zeta 
\left[{1\over F_\zeta}\big(\log F_\zeta\big)_{u\zeta}\right]_\zeta. 
 \label{Psi4}
 \end{equation} 
 These solutions always contain singular points on each wave surface, which
combine to form singular lines, but these are generally not well understood.

It can be seen from (\ref{Psi4}) that, in the case when $F$ is independent of $u$,
the above solution is conformally flat. This is just Minkowski, de~Sitter or
anti-de~Sitter space according to the value of $\Lambda$. Since we will consider
sandwich Robinson--Trautman waves below, we will refer to these as background
spaces and treat them separately in the next two sections. Although it is
obviously not the only possibility, these are the simplest conformally flat limits
of these solutions. They correspond to the natural choice of $f(\xi,u)=0$ in the
coordinates of \cite{GarPle81} (see also \cite{BicPod99a}).

\section{Coordinates in the Minkowski background}

Let us first consider the Minkowski metric in the form 
 \begin{equation}
 \d s^2 =2\,\d{\cal U}\,\d{\cal V}-2\,\d\eta\,\d\bar\eta,
 \label{MinkMetric}
 \end{equation} 
 in which the coordinates are related to the usual cartesian coordinates by 
 \begin{equation}
 {\cal U}={\textstyle{1\over\sqrt2}}(t+z), \qquad
 {\cal V}={\textstyle{1\over\sqrt2}}(t-z), \qquad
 \eta={\textstyle{1\over\sqrt2}}(x+iy). 
 \label{MinkNull}
 \end{equation} 
 Then the transformation
 \begin{equation}
 {\cal U}=u+{r\,F\bar F\over1+\epsilon F\bar F}, \qquad 
{\cal V}=\epsilon u+{r\over1+\epsilon F\bar F},
\qquad \eta={r\,F\over1+\epsilon F\bar F} ,
 \label{trans}
 \end{equation}
 where \ $F=F(\zeta)$ \ puts the Minkowski metric to the form
 $$ \d s^2 =2\,\d u\,\d r +2\epsilon\,\d u^2 -2{r^2\over P^2}\d\zeta\d\bar\zeta,
$$ 
 where $P$ is given by (\ref{P}). This is exactly the Minkowski limit of the
Robinson--Trautman family of solutions (\ref{RTmetric}) considered above in
which $\Lambda=0$ and $F$ is independent of $u$. The inverse transformation is
given by 
 \begin{eqnarray}
 u&=& {\cal U}-{1\over2\epsilon}\left[ \sqrt{({\cal V}-\epsilon\,U)^2 +4\epsilon
\eta\bar \eta} -({\cal V}-\epsilon\,{\cal U}) \right], \nonumber\\
 F(\zeta)&=& {1\over2\epsilon\bar\eta} \left[
\sqrt{({\cal V}-\epsilon\,{\cal U})^2 +4\epsilon \eta\bar \eta}
-({\cal V}-\epsilon\,{\cal U}) \right],  \label{Inverse}\\
 r&=& \sqrt{({\cal V}-\epsilon\,{\cal U})^2 +4\epsilon\eta\bar \eta}, \nonumber
 \end{eqnarray} 
 which reduces to \ $u={\cal U}-\eta\bar\eta/{\cal V}$, \ $F=\eta/{\cal V}$, \
$r={\cal V}$ \ in the limit when $\epsilon=0$.

From (\ref{Inverse}), it follows that the waves surfaces \ $u=u_0=$~const. \ are
given by 
 $$ \left(t-{\textstyle{1\over\sqrt2}}(1+\epsilon)u_0\right)^2 -x^2 -y^2 
-\left(z-{\textstyle{1\over\sqrt2}}(1-\epsilon)u_0\right)^2 =0. $$ 
 This describes a family of null cones. However, to assist with the physical
interpretation that will be presented later, we only consider the family of future
null cones relative to each vertex. These are clearly expanding spheres for all
values of $\epsilon$ and $u_0$. Successive cones are each displaced in the various
$t$ and $z$ directions as indicated in the space-time diagrams given in figure~1.

\begin{figure}[hpt]
\begin{center} 
\bigskip\bigskip
The eps version of this figure is too large to place on the archive. \break
A pdf version of this preprint with all figures is available at \break
www.lboro.ac.uk/departments/ma/preprints/papers02/02-27abs.html
\bigskip\bigskip
\caption{ Families of null cones given by \ $u=u_0$ \ foliate Minkowski
space-time in different ways in the three cases for which $\epsilon=1,0,-1$. (One
spatial dimension ($y$) is suppressed).   }
\end{center}
\end{figure}

When \ $\epsilon=+1$ \ all the null cones $u=$~const. naturally fit inside each
other and foliate the entire space-time. These surfaces are composed of expanding
concentric spheres. In this case, it can be seen from (\ref{MinkNull}) and
(\ref{Inverse}) that the origin of the Robinson--Trautman coordinate $r=0$
corresponds to $x=y=z=0$, with $t$ arbitrary, thus representing all the vertices
of the family of cones localised along the $t$ axis.

When \ $\epsilon=0$ \ the cones all have a common line $t=z$, $x=y=0$. The null
cones which represent expanding spheres, however, only foliate the half of the
Minkowski space given by $t>z$. (The other half, representing contracting
spheres, would cover the remaining half $t<z$.) In this case, the origin of the
Robinson--Trautman coordinate $r=0$ corresponds to the boundary plane $t=z$, with
$x$ and $y$ arbitrary.

When \ $\epsilon=-1$, \ however, the null cones all intersect each other. In this
case, the origin of the Robinson--Trautman coordinate $r=0$ corresponds to \
${t^2=x^2+y^2}$, \ with $z$ arbitrary. This clearly represents the envelope of
the above family of cones, and is a cylinder which expands at the speed of light.
The half null cones representing expanding spheres doubly foliate the inner
region $t\ge\sqrt{x^2+y^2}$. (Similarly, the half null cones representing
contracting spheres would doubly foliate the region $t\le-\sqrt{x^2+y^2}$). Points
for which $|t|<\sqrt{x^2+y^2}$ are all excluded in these foliations. For the
weak field limit of the Robinson--Trautman solutions, an overlapping of the wave
surfaces is not permitted. Thus each wave surface for this case must be
represented only by a family of half null cones for which $z\ge\sqrt2u_0$, as
indicated in figure~2. This restriction is necessary to maintain a consistent and
unambiguous foliation.

\begin{figure}[hpt]
\begin{center} 
\bigskip\bigskip
The eps version of this figure is too large to place on the archive. \break
A pdf version of this preprint with all figures is available at \break
www.lboro.ac.uk/departments/ma/preprints/papers02/02-27abs.html
\bigskip\bigskip
\caption{ Three typical members of the family of null cones $u=$~const. which
foliate part of Minkowski space in the case when $\epsilon=-1$. (One spatial
dimension ($y$) has been suppressed.)   }
\end{center}
\end{figure}

For the case $\epsilon=-1$, there appears to be another singularity when \
$P=0$, \ i.e. when \ $|F|=1$. \ It can be seen from (\ref{Inverse}) that this
occurs for \ ${t^2=x^2+y^2}$, \ with $z$ arbitrary. However, this exactly
coincides with $r=0$, which is the envelope of the null cones discussed above. It
can therefore be seen that the parts of null cones $t\ge0$, $z\ge\sqrt2u_0$,
which are the wave surfaces for this solution, are spanned by the above
coordinates with $|F|\le1$. (The other half of the cones $z\le\sqrt2u_0$ are
spanned by $|F|\ge1$.)

\section{Coordinates in the (anti-)de~Sitter background}

The above analysis of the character of the coordinates applies only to the
Minkowski background. However, a similar structure also occurs in de~Sitter and
anti-de~Sitter backgrounds. In these cases, however, it is the global structure
which is different.

It is well known that the (anti-)de~Sitter space can be represented as a
four-dimensional hyperboloid 
 \begin{equation}
 {Z_0}^2 -{Z_1}^2 -{Z_2}^2 -{Z_3}^2 -\varepsilon{Z_4}^2 =-\varepsilon a^2,
 \label{hyperboloid} 
 \end{equation}
 embedded in a five-dimensional Minkowski space-time
 $$ \d s^2= \d{Z_0}^2 -\d{Z_1}^2 -\d{Z_2}^2 -\d{Z_3}^2 -\varepsilon\d{Z_4}^2, $$ 
 where $a^2=3/|\Lambda|$ for a cosmological constant $\Lambda$,
$\varepsilon=1$ for a de~Sitter background ($\Lambda>0$), and $\varepsilon=-1$ for
an anti-de~Sitter background ($\Lambda<0$). Using the parameterization 
 \begin{equation}
 {\cal U}=\sqrt2a\,{Z_0+Z_1\over Z_4+a}, \qquad
{\cal V}=\sqrt2a\,{Z_0-Z_1\over Z_4+a}, \qquad
\eta=\sqrt2a\,{Z_2+iZ_3\over Z_4+a}, 
 \label{parameterization}
 \end{equation}
 the metric takes the form 
 \begin{equation}
 \d s^2 ={2\,\d{\cal U}\,\d{\cal V}-2\,\d\eta\,\d\bar\eta \over
\left[1+{\Lambda\over6}(\eta\bar\eta-{\cal UV})\right]^2},
 \end{equation} 
 which is explicitly conformal to Minkowski space (\ref{MinkMetric}) to which it
reduces when $\Lambda=0$. Applying now the transformation (\ref{trans}), the line
element becomes 
 $$ \d s^2 ={2\,\d u\,\d r +2\epsilon\,\d u^2 -2{r^2\over
P^2}\,\d\zeta\,\d\bar\zeta
\over \left[ 1-{\Lambda\over6}u(r+\epsilon u) \right]^2}. $$ 
 The further transformation 
 $$ \tilde r={r\over 1-{\Lambda\over6}u(r+\epsilon u)}, \qquad
 \tilde u=\int{\d u\over 1-\epsilon{\Lambda\over6}u^2}, $$ 
 yields 
 \begin{equation}
 \d s^2 =2\,\d\tilde u\,\d\tilde r
+\Big[2\epsilon -{\Lambda\over3}\tilde r^2\Big]\d\tilde u^2
-2{\tilde r^2\over P^2}\,\d\zeta\,\d\bar\zeta, 
 \end{equation} 
 which is clearly the (anti-)de~Sitter background expressed in the
Robinson--Trautman form (\ref{RTmetric}).

Let us now consider the wave surfaces $\tilde u=\tilde u_0=$~const. Clearly a
constant $\tilde u$ implies that $u$ is also a constant. Then, substituting
$u=u_0$ into (\ref{Inverse}) and the applying the parameterization
(\ref{parameterization}) gives 
 \begin{equation}
 (1+\epsilon){u_0\over\sqrt2 a}\,Z_0
-(1-\epsilon){u_0\over\sqrt2 a}\,Z_1
-\left(\varepsilon+\epsilon{{u_0}^2\over2a^2}\right)Z_4
+\left(\varepsilon-\epsilon{{u_0}^2\over2a^2}\right)a =0,
 \label{planes}
 \end{equation} 
 which is linear in $Z_i$, thus representing a family of planes in the
5-dimensional Minkowski space. Their sections through the 4-dimensional
hyperboloid (\ref{hyperboloid}) give the family of wave surfaces $\tilde
u=$~const. in the (anti-)de~Sitter universe. In fact, these planes are tangent to
the hyperboloid. In the following, it will be demonstrated that all the sections
are null cones relative to a vertex which is the point at which the plane touches
the hyperboloid. For this discussion, it will be convenient to introduce a
dimensionless parameter $\alpha$ such that \ $\tan({\alpha/2})
={u_0\over\sqrt2\,a}$, or 
 \begin{equation}
 \sin\alpha={2\,{u_0\over\sqrt2\,a} \over 1+{u_0^2\over2\,a^2}}, \qquad\qquad 
 \cos\alpha={1-{u_0^2\over2\,a^2} \over 1+{u_0^2\over2\,a^2}},
 \label{alpha}
 \end{equation}
 which will parameterize the family of wave surfaces.

\subsection*{The de~Sitter background}

For the case in which $\Lambda>0$ (i.e. $\varepsilon=1$), there are three cases to
consider.

We first investigate the case when $\epsilon=1$. Using (\ref{planes}) and the
parameter $\alpha$ given by (\ref{alpha}), the wave surfaces $u=u_0$ are given by
the planes \ $Z_0\sin\alpha-Z_4+a\cos\alpha=0$. \ Their intersections with the
hyperboloid (\ref{hyperboloid}) are 
 \begin{equation}
 (Z_0\cos\alpha-a\sin\alpha)^2=Z_1^2+Z_2^2+Z_3^2 ,
 \label{cones1a}
 \end{equation}
 which are a family of null cones with vertices localised on the timelike
hyperbola \ $Z_0=a\tan\alpha$, \ $Z_1=Z_2=Z_3=0$, \ $Z_4=a\sec\alpha$, \ which
corresponds to the origin of the Robinson--Trautman coordinates \ $\tilde r=0$. \
For this case, \ ${u_0\over\sqrt2\,a}=\tanh{\tilde u_0\over\sqrt2\,a}$. \ From
this it follows that \ $\tilde u_0\in(-\infty,\infty)$ \ corresponds to the
range \ $\alpha\in(-{\pi\over2},{\pi\over2})$. \ Thus, all the vertices are
located on an infinite timelike hyperbola with \ $Z_0\in(-\infty,\infty)$. \ The
future null cones from these vertices naturally foliate the half of the de~Sitter
space for which \ $Z_0+Z_4\ge0$. \ These are illustrated in figure~3a.

\begin{figure}[hpt]
\begin{center}
\bigskip\bigskip
The eps version of this figure is too large to place on the archive. \break
A pdf version of this preprint with all figures is available at \break
www.lboro.ac.uk/departments/ma/preprints/papers02/02-27abs.html
\bigskip\bigskip
\caption{ Families of null cones in de~Sitter space given by \ $u=u_0$ \ are
sections of a hyperboloid in a 5-dimensional Minkowski space. With two dimensions
($Z_2$ and $Z_3$) suppressed, they appear as straight (null) lines. These foliate
parts of the hyperboloid in different ways for the three cases in which
$\epsilon=1,0,-1$. The vertices are located respectively along timelike, null and
spacelike lines. }
\end{center}
\end{figure}

Next, we consider the case $\epsilon=0$, for which the wave surfaces are given by
the planes \ $Z_4-a={u_0\over\sqrt2\,a}(Z_0-Z_1)$. \ In this case, their
intersections with the hyperboloid are given by 
 \begin{equation}
 \left(Z_0-{\textstyle{u_0\over2\sqrt2\,a}}(Z_4+a)\right)^2
=\left(Z_1-{\textstyle{u_0\over2\sqrt2\,a}}(Z_4+a)\right)^2+Z_2^2+Z_3^2.
 \label{cones1b}
 \end{equation}
 Again, this is a family of null cones but with vertices now located along one
common straight null line \ $Z_0=Z_1={u_0\over\sqrt2}$, \ $Z_2=Z_3=0$, \ $Z_4=a$.
\ However, the origin of the Robinson--Trautman coordinates \ $\tilde r=0$ \
actually occurs on the null hypersurface \ $Z_0=Z_1$, \ with \
$Z_2^2+Z_3^2+Z_4^2=a^2$, \ which is a sphere with radius $a$ equal to that of the
universe at the instant $Z_0=0$. For this case, \ $u_0=\tilde
u_0\in(-\infty,\infty)$ \ covers the complete null line. The future null cones
from the vertices foliate the half of the de~Sitter space for which \
$Z_0-Z_1\ge0$. \ These are illustrated in figure~3b.

Finally, we consider the case when $\epsilon=-1$. Again, using (\ref{planes}) and
the parameter $\alpha$ from (\ref{alpha}), the wave surfaces $u=u_0$ are given by
the planes \ $Z_1\sin\alpha+Z_4\cos\alpha-a=0$. \ Their intersections with the
hyperboloid are 
 \begin{equation}
 Z_0^2= (Z_1\cos\alpha-Z_4\sin\alpha)^2+Z_2^2+Z_3^2,
 \label{cones1c}
 \end{equation}
 which are a family of null cones with vertices on the spacelike circle \
$Z_0=Z_2=Z_3=0$, \ $Z_1=a\sin\alpha$, \ $Z_4=a\cos\alpha$. \ These are
illustrated in figure~3c. For this case, \
${u_0\over\sqrt2\,a}=\tan{\tilde u_0\over\sqrt2\,a}$, \ so that \
$\alpha={\sqrt2\,\tilde u_0\over a}$. \ It is therefore necessary that \ $\tilde
u_0\in(-{a\pi\over\sqrt2},{a\pi\over\sqrt2}]$, \ or \ $\alpha\in(-\pi,\pi]$. \
Thus, the vertices are located on a closed circle around the de~Sitter universe
with \ $Z_0=Z_2=Z_3=0$. \ However, in this case, the future null cones from these
vertices would cover the future half \ ($Z_0\ge0$) \ of the de~Sitter space twice.
As for the Minkowski case with $\epsilon=-1$, again we must only consider the
family of half null cones with \ $Z_1\ge Z_4\tan\alpha$. \ Locally these are
similar to the wave surfaces illustrated in figure~2 but, in this case, they wrap
round the entire universe. Moreover, the origin \ $\tilde r=0$ \ now corresponds
to \ $Z_0^2=Z_2^2+Z_3^2$ \ with \ $Z_1^2+Z_4^2=a^2$. \ When $Z_0=0$, this is the
closed circle around the de~Sitter universe containing all the vertices of the
null cones. For $Z_0>0$, this a torus $S^1\times S^1$. Thus, in this case, the
origin of the Robinson--Trautman coordinates is a torus whose radius expands at
the speed of light in the de~Sitter background. We finally observe that the
apparent singularity which occurs when $P=0$ coincides exactly with this expanding
toroidal origin $\tilde r=0$.

\subsection*{The anti-de~Sitter background}

For the alternative case in which $\Lambda<0$ (i.e. $\varepsilon=-1$), there are
again three cases to consider.

First, when $\epsilon=1$, the wave surfaces $u=u_0$ are given by the planes \
$Z_0\sin\alpha+Z_4\cos\alpha-a=0$. \ Their intersections with the hyperboloid are 
 \begin{equation}
 (Z_0\cos\alpha-Z_4\sin\alpha)^2=Z_1^2+Z_2^2+Z_3^2,
 \label{cones2a}
 \end{equation}
 which are null cones with vertices on the closed timelike line \
$Z_0=a\sin\alpha$, \ $Z_1=Z_2=Z_3=0$, \ $Z_4=a\cos\alpha$ \ in the anti-de~Sitter
universe. This line corresponds to the origin of the Robinson--Trautman
coordinate \ $\tilde r=0$. \ Again in this case, \ $\alpha={\sqrt2\,\tilde
u_0\over a}$, \ so that \ $\tilde u_0\in(-{a\pi\over\sqrt2},{a\pi\over\sqrt2}]$ \
corresponds to \ $\alpha\in(-\pi,\pi]$. \ The future null cones from these
vertices, which are illustrated in figure~4a, now cover the complete
anti-de~Sitter space. It is also possible to consider the covering space for
which $\tilde u_0$ is permitted to take any value.

\begin{figure}[hpt]
\begin{center}
\bigskip\bigskip
The eps version of this figure is too large to place on the archive. \break
A pdf version of this preprint with all figures is available at \break
www.lboro.ac.uk/departments/ma/preprints/papers02/02-27abs.html
\bigskip\bigskip
\caption{ Families of null cones \ $u=u_0$ \ in anti-de~Sitter space are sections
of a hyperboloid in a 5-dimensional Minkowski space. With the spatial dimensions
$Z_2$ and $Z_3$ suppressed, they are straight (null) lines. These foliate parts
of the hyperboloid in different ways for the three cases in which
$\epsilon=1,0,-1$, with vertices located respectively along timelike, null and
spacelike lines.  }
\end{center}
\end{figure}

Next, when $\epsilon=0$, the wave surfaces are given by the planes \
$Z_4-a=-{u_0\over\sqrt2\,a}(Z_0-Z_1)$. \ In this case, their intersections with
the hyperboloid are the null cones 
 \begin{equation}
 \left(Z_0-{\textstyle{u_0\over2\sqrt2\,a}}(Z_4+a)\right)^2
=\left(Z_1-{\textstyle{u_0\over2\sqrt2\,a}}(Z_4+a)\right)^2+Z_2^2+Z_3^2.
 \label{cones2b}
 \end{equation}
 with vertices located on one common null line \ $Z_0=Z_1={u_0\over\sqrt2}$, \
$Z_2=Z_3=0$, \ $Z_4=a$. \ However, the origin of the Robinson--Trautman
coordinate \ $\tilde r=0$ \ actually occurs on the null hypersurface \ $Z_0=Z_1$
\ with \ $Z_4^2=a^2+Z_2^2+Z_3^2$, \ which is a hyperboloidal surface. With \
$u_0=\tilde u_0\in(-\infty,\infty)$, \ the future null cones foliate the half of
the anti-de~Sitter space for which \ $Z_0-Z_1\ge0$. \ These are illustrated in
figure~4b.

Finally, when $\epsilon=-1$, the wave surfaces $u=u_0$ are given by the planes \
$Z_1\sin\alpha-Z_4+a\cos\alpha=0$, \ which intersect the hyperboloid on the null
cones
 \begin{equation}
 Z_0^2= (Z_1\cos\alpha-a\sin\alpha)^2+Z_2^2+Z_3^2.
 \label{cones2c}
 \end{equation}
 These are illustrated in figure~4c. Their vertices are located on the spacelike
hyperbola \ $Z_0=Z_2=Z_3=0$, \ $Z_1=a\tan\alpha$, \ $Z_4=a\sec\alpha$. \  For
this case, \ ${u_0\over\sqrt2\,a}=\tanh{\tilde u_0\over\sqrt2\,a}$. \ Here again \
$\tilde u_0\in(-\infty,\infty)$ \ corresponds to \
$\alpha\in(-{\pi\over2},{\pi\over2})$, \ so that \ $Z_1\in(-\infty,\infty)$. \ As
for the previous cases in which $\epsilon=-1$, to provide an unambiguous
foliation, we must again only consider the family of half null cones with \
$Z_1\ge a\tan\alpha$, \ which are analogous to the wave surfaces illustrated in
figure~2. Moreover, the origin \ $\tilde r=0$ \ of the Robinson--Trautman
coordinates corresponds to \ $Z_0^2=Z_2^2+Z_3^2$ \ with \ $Z_4^2=a^2+Z_1^2$. \
When $Z_0=0$, this is a hyperbola across the anti-de~Sitter universe which
contains all the vertices of the null cones. For $Z_0>0$, it is an expanding
cylindrical-type surface $S^1\times H^1$ centred on this hyperbola. We again
observe that the singularity $P=0$ coincides exactly with $\tilde r=0$.

\section{A class of Robinson--Trautman solutions}

The family of all type N Robinson--Trautman solutions (\ref{RTmetric}) are
characterised by two parameters $\Lambda$ and $\epsilon$, and an arbitrary
holomorphic function $F(\zeta,u)$. To investigate these space-times and their
physical interpretation, it is convenient to restrict attention to the
particularly simple case in which 
 \begin{equation}
 F(\zeta,u)=\zeta^{g(u)}, 
 \label{F}
 \end{equation}
 where $g(u)$ is an arbitrary positive function of retarded time. For this
choice, the expressions involving $P$ given by (\ref{P}), which appear in the
metric (\ref{RTmetric}), take the forms 
  \begin{equation}
 P^2={\left[1+\epsilon(\zeta\bar\zeta)^g\right]^2
\over g^2(\zeta\bar\zeta)^{g-1}}\ , \qquad 
 (\log P)_u =-{\>g'\over g} \left( 1 +{\textstyle{1\over2}}
\log(\zeta\bar\zeta)^g 
\left[ {1-\epsilon(\zeta\bar\zeta)^g\over1+\epsilon(\zeta\bar\zeta)^g} \right]
  \right),
  \label{functions}
  \end{equation}
 and the explicit expression for $\Psi_4$ given by (\ref{Psi4}) is 
  \begin{equation}
 \bar\Psi_4=-{g'\bar\zeta\over 2g\,\zeta\,r}
\left({1\over|\zeta|^{g}}+\epsilon|\zeta|^{g}\right)^2\ .
  \label{Psi4case}
  \end{equation}
Obviously, the solution is conformally flat (i.e. the Minkowski, de Sitter or
anti-de Sitter backgrounds) if and only if $g$ is a constant. In general it
represents an exact gravitational wave with arbitrary amplitude. It reduces to a
weak radiation field if $g$ is approximately constant, i.e. when $g'/g$ is
small. Notice that, interestingly, the term $(\log P)_u$ which appears in the
metric and the Weyl tensor component $\Psi_4$ are both proportional to the same
wave profile $g'/g$.

Since $\Psi_4$ is the only non-vanishing component of the Weyl tensor, it is
clear that all the familiar scalar polynomial invariants vanish. These solutions
therefore do not contain scalar polynomial curvature singularities. However, a
second order invariant for expanding type~N space-times has been found by
Bi\v{c}\'ak and Pravda \cite{BicPra98}. This is given by 
 $$ I=(48\,\rho^2\bar\rho^2\Psi_4\bar\Psi_4)^2, $$ 
 where $\rho=1/r$ in this case. It follows that the above solutions possess
curvature singularities when $r=0$ and when $\zeta=0$ or $\infty$, provided
$g'\ne0$. Some aspects of these singularities will be discussed in detail below.

\section{Sandwich Robinson--Trautman waves}

A particular sandwich Robinson--Trautman wave of the above form has recently been
presented in \cite{GriDoc02} for the case $\Lambda=0$ and $\epsilon=1$. This has
the form (\ref{F}) with 
 \begin{equation}
 g(u) = \left\{ \begin{array}{lll}
 1-ab &{\rm for} \ u<0 \\
 \noalign{\smallskip}
 1-a(b-u)
 \qquad &{\rm for} \ 0\le u\le b \\
 \noalign{\smallskip}
 1 &{\rm for} \ u>b
\end{array} \right.
 \label{LettSoln}
 \end{equation}
 where $a$ and $b$ are positive constants. This solution represents a
Robinson--Trautman wave confined to the region \ $0\le u\le b$, \ in which $g'=a$.
It has an expanding spherical wavefront $u=0$, and the wave continues until $u=b$,
which is also a concentric expanding sphere. Ahead of and behind this wave zone,
the space-time is the locally flat Minkowski background. However, taking \
$\arg\zeta\in[0,2\pi)$, \ the Minkowski region ahead of the wave contains a
topological defect localised at $\zeta=0$ and $\zeta=\infty$. This can be
interpreted as a cosmic string with the deficit angle $2\pi ab$. By contrast, the
Minkowski region behind the wave contains no such defect. This solution has thus
been interpreted as representing a snapping cosmic string in a Minkowski
background in which the tension (deficit angle) of the string reduces uniformly
to zero over a finite interval of retarded time. This decay of the cosmic string
may be considered to generate the gravitational wave.

The above solution can be generalised to arbitrary functions $g(u)$, and to all
possible values of $\Lambda$ and $\epsilon$. Obviously, $g(u)$ specifies the wave
profile, $\Lambda$ determines the background, and $\epsilon$ characterises the
geometrical properties of the wave surfaces.

First, it can be seen that on any wave surface $u=$~const., the complex number
$F=\zeta^g$ represents a stereographic-type coordinate. Introducing the
parameterisations, 
 \begin{equation}
 \zeta^g = \left\{ \begin{array}{lll}
 \cot({\textstyle{1\over2}}\theta)\,e^{i\phi} &{\rm for} \ \epsilon=1 \\
 \noalign{\smallskip}
 {\textstyle{1\over2}}\varrho\,e^{i\phi}
 \qquad &{\rm for} \ \epsilon=0 \\
 \noalign{\smallskip}
 \tanh({\textstyle{1\over2}}R)\,e^{i\phi} \qquad &{\rm for} \ \epsilon=-1
\end{array} \right.
 \end{equation}
 the wave surfaces take the standard form 
 \begin{equation}
 {4\over P^2}\,\d\zeta\,\d\bar\zeta = \left\{ \begin{array}{lll}
 \d\theta^2+\sin^2\theta\>\d\phi^2 &{\rm for} \ \epsilon=1 \\
 \noalign{\smallskip}
 \d\varrho^2+\varrho^2\>\d\phi^2
 \qquad &{\rm for} \ \epsilon=0 \\
 \noalign{\smallskip}
 \d R^2+\sinh^2 R\>\d\phi^2 \qquad &{\rm for} \ \epsilon=-1
\end{array} \right.
 \end{equation}
 This is simply the metric on a two-sphere, two-plane, and two-hyperboloid,
respectively. (Note that for the case $\epsilon=-1$, we restrict $\zeta$ to the
range $|\zeta|\le1$ to cover a single sheet of the hyperboloid.) However, if we
assume that the argument of $\zeta$ covers the full range $[0,2\pi)$, then
$\phi\in[0,2\pi g)$ and these surfaces in general include a deficit angle
$2\pi(1-g)$ around $\theta=0$ or $\pi$ if $\epsilon=1$, around $\varrho=0$ if
$\epsilon=0$, and around $R=0$ if $\epsilon=-1$.

Clearly, any region in which $g=$~const. must be one of the Minkowski, de~Sitter
or anti-de~Sitter backgrounds according to the value of $\Lambda$. In cases with
$g<1$, these regions contain a constant deficit angle on all sections
$u=$~const. Together, these are interpreted as forming cosmic strings with
constant tension. (If $g>1$, there is an excess angle corresponding to a strut of
constant compression.) When $g=1$ there is no string.

In general, in any region in which $g(u)$ is not constant, this is a
Robinson--Trautman type~N solution, and the deficit angle on each wave surface
will vary. However, it can be seen from (\ref{Psi4case}) that the poles about
which there is a deficit angle actually correspond to curvature singularities in
the Weyl tensor.

By combining the above two possibilities, we can construct solutions in
which $g$ is non-constant only over a {\it finite} range of~$u$. Such solutions
clearly represent sandwich Robinson--Trautman waves. The situation in which $g(u)$
is constant ($<1$) in front of the wave and then increases continuously to 1
behind the wave can be interpreted as a disintegrating string (the deficit angle
reduces continuously to zero). One particular case of this, described in
\cite{GriDoc02} for $\epsilon=1$, is represented by (\ref{LettSoln}) and
illustrated in figure~5a. The analogous solutions for alternative values of
$\epsilon$ are also illustrated in figure~5.

\begin{figure}[hpt]
\begin{center} \includegraphics[scale=0.6, trim=5 30 5 -5]{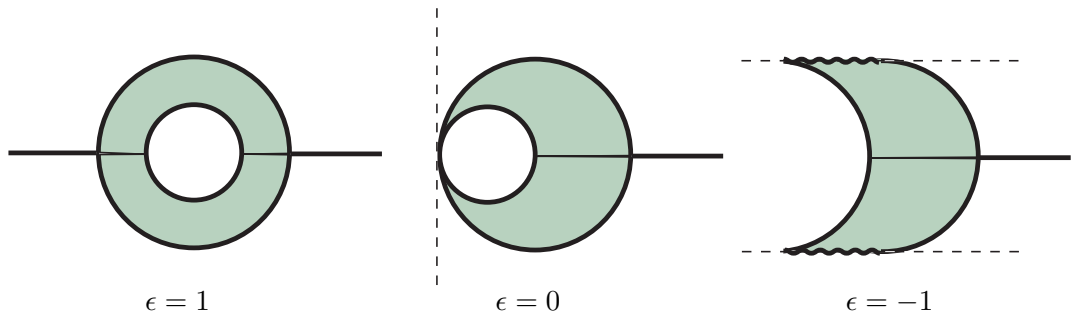}
 $$ \epsilon=1 \hskip9pc \epsilon=0 \hskip9pc \epsilon=-1 $$
 \vskip-1pc
\caption{ The shaded regions represent Robinson--Trautman sandwich waves at some
fixed time for different values of $\epsilon$. The expanding spherical
(hemispherical for $\epsilon=-1$) wave surfaces are given by $u =$ const. The
region behind the wave is Minkowski or (anti-)de~Sitter, while the region ahead of
the wave is Minkowski or (anti-)de~Sitter with a deficit angle representing a
cosmic string. The dashed lines denote the boundaries of the coordinate system
adopted. }
\end{center}
\end{figure}

As explained in sections 3 and 4, when $\epsilon=1$ the wave surfaces $u=$~const.
at any time are a family of concentric spheres. For the above example, the
background space ahead of the wave contains strings of equal tension at opposite
sides of the expanding spherical wavefront at $\theta=0$ and $\theta=\pi$. 

In the case when $\epsilon=0$, there is only a single pole in $\varrho=0$. The
background region thus contains a single string, but only part of the complete
space-time is now covered by the coordinate system (that to the right of the
dashed line in figure~5b). At any time, the spherical wave surfaces contain a
common point, opposite to the pole, as may be observed from the family of null
cones in figure~1b. 

When $\epsilon=-1$, since we have made the restriction $|\zeta|\le1$ here to span
only hemispherical surfaces, there is only a single pole $R=0$ on each expanding
hemisphere. This may again be attached to a cosmic string in the background
region ahead of the wavefront. As explained in previous sections, the envelope of
these surfaces $r=0$ is a physical singularity within the wave zone. In the
background space-time, it is a cylindrical surface of finite length whose radius
is expanding at the speed of light. Its section is denoted by the two thick wavy
line in figure~5c. The boundary of the coordinate system is again illustrated in
figure~5c by the dashed line.

For the particular choice of the sandwich wave (\ref{LettSoln}), the function
$g(u)$ is linear within the wave zone. This gives rise to discontinuities in
$\Psi_4\sim g'(u)$ at $u=0$ and $u=b$, see (\ref{Psi4case}), which correspond to
shocks on the boundary wave surfaces indicated in figure~5. However, for this
example, the $(\log P)_u$ term contained in the metric is also discontinuous on
these shock fronts, see (\ref{functions}). (It may be noted that shock
Robinson--Trautman waves which have a continuous metric form have been obtained by
Nutku \cite{Nutku91}, but these are not considered further here.) Of course, more
general families of sandwich waves without discontinuities in the metric and
$\Psi_4$ can easily be constructed by permitting $g$ to vary arbitrarily over
only a finite range of the retarded time. In this way, we can obtain explicit
solutions in which $\Psi_4$ may be an arbitrarily smooth function of $u$.  In
addition, it is particularly interesting to note that, if $g=1$ on both sides of
the sandwich, the background does not contain a cosmic string either in front of
or behind the wave.

\section{The character of singularities and global structure}

The illustrations in figure~5 are useful for identifying the location of cosmic
strings in the outer region and of string-like structures which are their
continuations in the wave zones. However, as can be seen from (\ref{Psi4case}),
with $g'\ne0$, these string-like structures at $\zeta=0$ (and $\zeta=\infty$ for
$\epsilon=1$) in the wave regions are space-time curvature singularities. These
singularities propagate at the speed of light along with the sandwich wave.

In addition, these space-times also always contain singularities when $r=0$. These
may naturally be considered as the sources of the Robinson--Trautman solutions.
The character of the coordinate origin $r=0$ in the background space-times for
different values of $\epsilon$ and $\Lambda$ has already been discussed in
sections 3 and 4. For sandwich Robinson--Trautman waves, this curvature
singularity occurs only within the wave zones, and does not extend to the regions
in front of and behind the waves where $g'=0$. The location of this singularity
for sections of constant~$\zeta$ and different values of $\epsilon$ are
illustrated in the space-time pictures in figure~6.

\begin{figure}[hpt]
\begin{center} \includegraphics[scale=0.6, trim=5 5 5 -5]{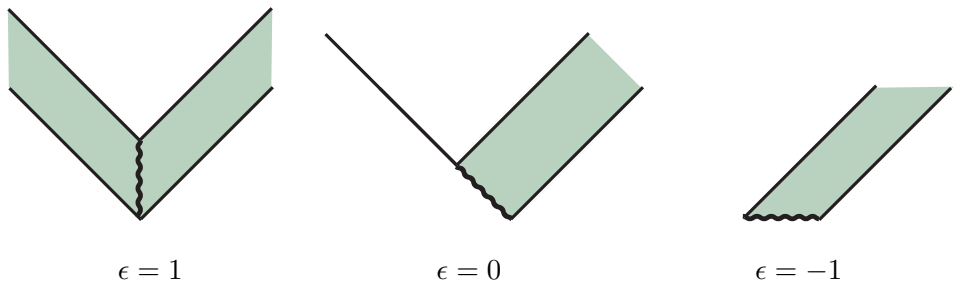}
 $$ \epsilon=1 \hskip8pc \epsilon=0 \hskip8pc \epsilon=-1 $$
 \vskip-1pc
\caption{ Schematic space-time pictures for constant $\zeta$ for
Robinson--Trautman sandwich waves with different values of~$\epsilon$. The shaded
areas are the wave regions. The thick wavy lines represent the singularities at
$r=0$. If the particular section corresponds to $\zeta=0$, the region external to
the wave also contains a cosmic string which is extended as a string-like
singularity through the wave zone.  }
\end{center}
\end{figure}

It should be noted that figure~5 corresponds to a spacelike section through the
space-time pictures in figure~6 with the addition of one spatial dimension. 
The particular section in figure~5 is at a sufficiently late time to contain the
complete sandwich. Earlier spacelike sections than those shown in figure~5 could
intersect the naked curvature singularity at $r=0$, which corresponds to the
source of the wave.

For the cases in which $\epsilon=1$ and $\epsilon=0$, the curvature singularity at
$r=0$ corresponds to a timelike or null {\it line} respectively, at least in the
weak field limit. However, for the case in which $\epsilon=-1$, at an initial time
this is a spacelike line of finite length. This line subsequently becomes a
cylinder whose radius expands at the speed of light. For a sandwich wave, this
expanding cylinder has constant finite length.

For the case in which $\epsilon=1$, the solution can be interpreted as
representing the snapping of a cosmic string in which the deficit angle of the
string reduces to zero through a region in which it generates the gravitational
wave.

A similar interpretation can be given for the case when $\epsilon=0$. However,
there is now a boundary of the coordinate systems which is represented as a dashed
line in figure~5b. This is a {\it null plane} in the background and, as such, it
is possible to locally extend the space-time to include an external conformally
flat background region. Such an extension is non-unique as an arbitrary impulsive
gravitational wave may occur on this null hypersurface. In the absence of such an
impulsive wave, it is necessary to have the same backgrounds which contain strings
in both directions: i.e. the region to the left of the dashed line in figure~5b is
taken to be the same as the region external to the wavefront, having a string with
the same deficit angle. In this case, one end of the string is a clean break,
while the deficit angle at the other end reduces to zero over a finite range.

The situation when $\epsilon=-1$ is even more complicated. However, a model for a
snapping and disintegrating cosmic string can again be obtained by considering
two separate sandwich waves of the type described above and illustrated in
figures 5c and 6c. These can be combined in such a way that the hemispherical
waves propagate in opposite directions as illustrated in figure~7. This enables
us to construct a solution in which an initial cosmic string, at an initial time
and for some finite section, becomes a curvature singularity. This cylindrical
singularity subsequently expands radially at the speed of light (perpendicular to
the string) while maintaining constant length. This generates expanding
hemispherical gravitational waves at both ends. The deficit angle of the string
may reduce to zero through the finite sandwich wave regions in both directions.
This would leave an expanding spherical region of the background space-time
between the ends of the strings and inside the expanding singular cylinder. This
construction, however, requires at least three separate coordinate patches: one
representing each of the two sandwich waves, and one to represent the background
region external to the expanding cylinder (exterior to the dashed lines in
figure~5c).

\begin{figure}[hpt]
\begin{center} \includegraphics[scale=0.6, trim=5 5 5 -5]{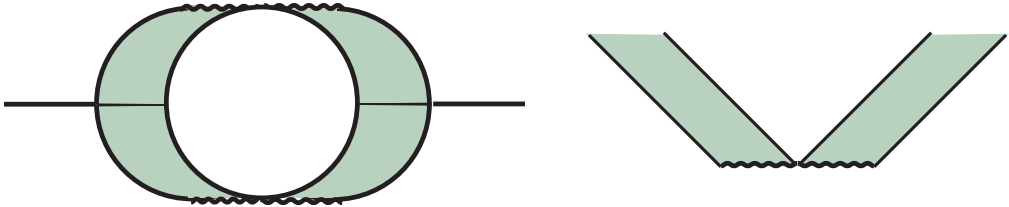}
\caption{ A snapping cosmic string in the case $\epsilon=-1$. The left picture
represents the situation at a fixed time (one spatial direction is suppressed).
The right picture is a space-time diagram with two spatial directions suppressed. 
The shaded areas are the wave regions. The thick wavy lines represent the
expanding cylindrical singularity where $r=0$.  }
\end{center}
\end{figure}

The properties described above can all be naturally interpreted in a Minkowski
background for the case in which $\Lambda=0$. When $\Lambda\ne0$, the geometry of
the sandwich waves illustrated in figure~5 remains basically the same. It is only
the geometry of the background which significantly differs. In particular, the
snapping straight string in a Minkowski background is replaced by a closed string
around the de~Sitter universe which snaps at a single event, or an infinite
snapping string in an anti-de~Sitter background. The structures of the
singularity $r=0$ also naturally carry over when $\Lambda\ne0$ if $\epsilon=1$ or
$\epsilon=0$. For $\epsilon=-1$ the expanding cylindrical section is replaced by
an expanding section of a torus $S^1\times S^1$ for $\Lambda>0$, and an expanding
section of the surface $S^1\times H^1$ for
$\Lambda<0$.

The pictures in figure~6 also remain valid for any value of $\Lambda$, and it is
possible to use these as parts of complete conformal diagrams. However, these
would involve including conformal boundaries representing null infinity which
has a null, spacelike or timelike character according as $\Lambda=0$, $\Lambda>0$
or $\Lambda<0$ respectively.

\section{Further observations}

We have presented an explicit family of Robinson--Trautman type~N solutions
$RTN(\Lambda,\epsilon)$ given by (\ref{F}). This has proved to be very suitable
for the physical interpretation of these solutions. In particular, it has enabled
us to construct sandwich waves of this class and to analyse their geometrical
properties and the character of the singularities. These all have (hemi-)spherical
wave surfaces and propagate in Minkowski, de~Sitter or anti-de~Sitter
backgrounds.

For all the sandwich waves described above, it is straightforward to construct
their corresponding impulsive limits. These are obviously expanding spherical
impulsive waves which are well known in Minkowski and (anti-)de~Sitter backgrounds
\cite{Penrose72}--\cite{PodGri01}. It has been argued \cite{Penrose72},
\cite{PodGri99a} that these are impulsive limits of Robinson--Trautman type~N
solutions, but this limit can here be performed explicitly, without incurring any
of the expected difficulties of dealing with the square of a Dirac delta
function in the metric.

To construct an impulse on the wave surface $u=u_0$, the value of $g$ is simply
taken to contain a step function $\Theta(u-u_0)$, so that
$\Psi_4\sim\delta(u-u_0)$. This construction necessarily gives rise to deficit
angles in one region or the other, so that such solutions can be interpreted as
snapping or expanding cosmic strings. Moreover, it is possible to consider two
(or more) sandwich waves, each in their impulsive limits, so that
$g\to1+a\Theta(u-u_0)-a\Theta(u-u_1)$, such that the background ahead of the
first impulse and behind the second impulse contain no strings (an outline of
this has been proposed in \cite{NutPen92}). In this sense, a pair of string
sections appear and propagate in opposite directions at the speed of light.

The initial ansatz (\ref{F}) may also be generalised. For example putting
 $$ F(\zeta,u)=\Big(\zeta-h(u)\Big)^{g(u)}, $$ 
 where $h(u)$ is a complex function, gives rise to the Weyl tensor component
 $$ \bar\Psi_4=-{1\over2r} {(\bar\zeta-\bar h)\over(\zeta-h)}
\left({g'\over g} +h'{\,1-g^{-2}\over\zeta-h} \right)
\left({1\over|\zeta-h|^{g}}+\epsilon|\zeta-h|^{g}\right)^2, $$ 
 which reduces to (\ref{Psi4case}) when $h=0$. This indicates that the singular
point on the spherical wave surfaces varies continuously from one surface to
another. This would cause the string-like singularity to bend within the wave
region. Other functions could introduce additional singular points so that, for
example, string-like singularities could even bifurcate. Such behaviour of course
depends on the specific character of the holomorphic function $F(\zeta,u)$.

The above solutions, however, are obviously not complete since they do not
contain their own past. They only describe the future of events like snapping and
disintegrating strings without providing any reasons for those strings to snap
or decay.

A remarkable feature of the Robinson--Trautman family of solutions is that they
place no restrictions at all on the character and decay of string-like structures
of the type described above. Any form of snapping or decaying string could be
accommodated within this framework.

\section*{Acknowledgements}

This work was supported in part, by the grant GACR-202/02/0735 of the Czech
Republic and GAUK~141/2000 of Charles University.

\vfill
\noindent
This paper is due to appear in {\sl Classical and Quantum Gravity} {\bf 19}
(2002). \hfil\break
It is also available, with all figures included, in pdf format at

www.lboro.ac.uk/departments/ma/preprints/papers02/02-27abs.html

\end{document}